\documentclass[12pt]{article}
\usepackage{amsfonts}
\usepackage{cite}

\begin{document}

\begin{center}
{\LARGE{On spacetime algebra and its relations with negative masses}}
\end{center}

\begin{center}
N. Debergh $^1$ and J.-P. Petit $^2$  
\vspace{5mm}\\
{\footnotesize{$^1$ Department of pedagogy, Haute Ecole Charlemagne, 1, rue Gr\'egoire Bodart, 4500 Huy, Belgium, nathalie.debergh@hech.be\vspace{2mm}}} \\
{\footnotesize{$^2$ Former Director of Research in Centre National de la Recherche Scientifique (C.N.R.S.),B.P.55, 84089 Pertuis, France, jppetit1937@yahoo.fr \vspace{2mm}}}
\end{center}

\begin{abstract}
We consider four subsets of the complexified spacetime algebra, namely the real even part, the real odd part, the imaginary even part and the imaginary odd part. This naturally leads to the four connected components of the Lorentz group, supplemented each time by an additional symmetry. We then examine how these four parts impact the Dirac equation and show that four types of matter arise with positive and negative masses as well as positive and negative charges.
\end{abstract}


\section{Introduction}
Geometric algebra was discovered jointly by Clifford and Grassmann in the late 19th century \cite{Grassmann, Clifford}. Despite its innovative and promising aspects, this algebra did not get the attention it deserved at that time. In fact, most researchers have only exploited the purely algebraic aspects of this concept and have obscured the other facets.
\par
It was not until the work of Hestenes \cite{Hestenes1966} that the richness of geometric algebra was fully realized. Since then, some scientists have made significant contributions and firmly believe (an opinion we share) that geometric algebra is the simplest and most coherent language that can reconcile physics with mathematics. The subject itself is covered in detail in \cite{Hestenes1966, Hestenes-Sobczyk1984, DL2003}.\par
Geometric algebra can be considered in any physical discipline but where it probably takes a particular importance is the physics of the electron \cite{DLGSC1996}. Generalizing the concept of complex numbers, it allows to highlight an interesting alternative to the Dirac theory. This other option is referred to as the real formulation of the Dirac equation or the Dirac-Hestenes equation \cite{Hestenes1967, Hestenes1975, DLG1993}.\par
This proposition is based on the geometric algebra related to the Minkowski spacetime (namely the spacetime algebra) and, surprisingly enough, only one part of this algebra (the part referred to as the real even part) is necessary to have equivalence with the Dirac formulation.\par
On a different level, the acceleration of the expansion of the Universe has been observed by astronomers \cite{Riess1998}. This discovery implies the existence of unknown matter and energy (i.e. the dark matter and the dark energy). If their natures are still a mystery, there is more and more evidence to say that particles of negative masses \cite{Bondi1957} could be the best candidates to solve this enigma \cite{BP2013, PD2014, MP2014, Hammond2015, Dvoeglazov2016, PD2018, Guay2022}.\par
The negative mass matter interacting only by gravity with positive mass matter (and not through electromagnetic phenomena), it is not possible to directly experiment it. Its detection is done indirectly by analysing the alteration of the light emitted by positive mass matter.  We refer the interested reader to \cite{PDD2021} for further details.\par
It is natural to wonder about a possible relation between spacetime algebra and negative masses and it is precisely the goal of this paper.\par
In Section 1, we will show, after a brief recall on the basics of the geometric algebra, that if only a part of the content (and by this we mean the even real part) of the spacetime algebra has been really exploited in most of the contributions on the geometric algebra of the electron, it is also possible to consider the other parts (odd real part, even imaginary part and odd imaginary part). These supplementary parts are connected to the usual one by applying to it the discrete unitary operators $P$,$T$,$PT$.\par
This will naturally leads us, in Section 2, to revise the relativistic invariance where not only the restricted Lorentz group ${\rm SO}^+ (1,3)$ will be reached but also its three connected components. When the odd part is included, we will show that the Lie superalgebra ${\rm gl}(2|2)$ arises naturally.\par
Section 3 will be devoted to the highlighting of the four types of solutions of the Dirac equation, both with positive and negative masses. \par
In Section 4, we will put in evidence the corresponding equations and solutions within the geometric algebra formalism.\par
The relativistic invariance of these equations as well as the transformation laws on their solutions will be discussed in Section 5. The remarkable point is the appearance of an additional symmetry compared to the traditional theory, symmetry which brings a transformation law on the mass.\par
We will conclude in the last Section.

\section{Preliminary: the spacetime algebra and its four components}
First, we briefly recall the fundamentals of a geometric algebra. \\
Let $u$ and $v$ be two vectors of a vector space of dimension $n$
\begin{equation}
u=u^j e_j,  \quad  v=v^j e_j,
\label{eq: 1}
\end{equation}
(here and in the following, summation on repeated indices is understood).\\
The geometric product of these vectors is defined according to
\begin{equation}
u \; v=u . v + u \wedge v.
\label{eq: 2}
\end{equation}
In the first term, we recognize the -symmetric- inner product (equivalent to the scalar product when dealing exclusively with vectors). It gives rise to a scalar ($0$-vector). In the second term, we find the -antisymmetric- outer product leading to a bivector ($2$-vector or oriented portion of plan).\\
In other words
\begin{equation}
u . v = \frac{1}{2} (u \; v+v \; u), \quad u \wedge v = \frac{1}{2} (u \; v-v \; u).
\label{eq: 3}
\end{equation}
Applied to the (orthogonal) vector basis, the geometric product is such that
\begin{equation}
e_j \; e_k =- e_k \; e_j, \quad j \neq k.
\label{eq: 4}
\end{equation}
It also generates a $2^n$-dimensional algebra made of one $0$-vector, n $1$-vectors, $\frac{n(n-1)}{2}$ $2$-vectors and so on until one $n$-vector, respectively given by
\begin{equation}
1, \quad e_j, \quad e_j \; e_k, \quad..., \quad I_n \equiv e_1 \; e_2 \; ...\; e_n.
\label{eq: 5}
\end{equation}
This $n$-vector $I_n$ is also known as the pseudo-scalar unit.\\
Second, we particularize to spacetime algebra (STA) for which $n=4$ and
\begin{equation}
e_0^2 =-e_1^2 =-e_2^2=-e_3^2=1. 
\label{eq: 6}
\end{equation}
From Eqs.~(4) and (6), it is obvious to see that the basis vectors of the STA are nothing but the generators of the Clifford algebra ${\rm Cl}(1,3)$ and can be realized through the Dirac matrices (in the Weyl realization, for instance):
\begin{equation}
e_0 = \gamma_0=\left(
\begin{array}{cc}
0 & I \\
I & 0
\end{array}
\right) \quad , \quad e_j=\gamma_j=\left(
\begin{array}{cc}
0 & \sigma_j \\
-\sigma_j & 0
\end{array}
\right), 
\label{eq: 7}
\end{equation}
where $I$ is the identity matrix of dimension 2 and $\sigma_j$ are the Pauli matrices.\\
A remarkable fact about STA is that its real even part
\begin{equation}
E=a^0 \; I+a^j \; \gamma_j \gamma_0-\frac{1}{2} \; \epsilon_j^{kl} \; b^j \; \gamma_k \gamma_l+b^0 \; I_4, \quad a^0, a^j, b^0, b^j \in \mathbf{R},
\label{eq: 8}
\end{equation}
(where $\epsilon_j^{kl}$ is the completely antisymmetric Levi-Civita tensor) is a geometric algebra in itself, namely the one associated with the 3D Euclidian space (SA). This is easily seen by identifying the first three $2$-vectors of the STA with the $1$-vectors of the SA
\begin{equation}
e'_j \equiv \gamma_j \gamma_0.
\label{eq: 9}
\end{equation}
It is then straightforward to see that
\begin{equation}
e'_j e'_k =-\gamma_j \gamma_k, \; I_3=e'_1 e'_2 e'_3=\gamma_0 \gamma_1 \gamma_2 \gamma_3 = I_4.
\label{eq: 10}
\end{equation}
Moreover, we can write
\begin{equation}
E=a^0 \; I+a^j \; e'_j +I_4 (b^0 \; I+b^j \; e'_j)=a^0 \; I+\vec a+I_4 (b^0 \; I+\vec b)=a+I_4 \; b.
\label{eq: 11}
\end{equation}
Here and in the following, we refer to the concept of relative vector \cite{Hestenes1974} (written with a lowercase letter)
\begin{equation}
u \equiv u^0 \; I+u^j \; e'_j =u^0 \; I+ \vec u,
\label{eq: 12}
\end{equation} 
by opposition to the one of proper vector (written with a capital letter)
\begin{equation}
U \equiv u \; e_0=u^{\mu} \; e_{\mu}, \quad \mu=0, 1, 2, 3.
\label{eq: 13}
\end{equation}
In addition to this real even part, we will consider three other contributions. In total, we will have four types of elements of the (complexified) STA:
\begin{itemize}
\item the real even part
$$
E=a+I_4 \; b,
\label{eq: 14a}
\eqno{(14a)}
$$
\item the real odd part
$$
O \equiv \gamma_0 \; E=a^0 \; \gamma_0 - a^j \; \gamma_j - \frac{1}{2} \; \epsilon_j^{kl} \; b^j \; \gamma_0 \gamma_k \gamma_l +b^0 \; \gamma_1 \gamma_2 \gamma_3,
\label{eq: 14b}
\eqno{(14b)}
$$
\item the imaginary even part
$$
IE \equiv -i \; I_4 \; E=i \; b^0 \; I-i \; b^j \; \gamma_0 \gamma_j + \frac{i}{2} \; \epsilon_j^{kl} \; a^j \; \gamma_j \gamma_k - i \; a^0 I_4,
\label{eq: 14c}
\eqno{(14c)}
$$
\item the imaginary odd part
$$
IO \equiv -i \; \gamma_1 \gamma_2 \gamma_3 \; E=i \; b^0 \gamma_0 - i \; b^j \; \gamma_j + \frac{i}{2} \; \epsilon_j^{kl} \; a^j \; \gamma_0 \gamma_k \gamma_l -i \; a^0 \; \gamma_1 \gamma_2 \gamma_3.
\label{eq: 14d}
\eqno{(14d)}
$$
\end{itemize}
Notice that the great majority of the authors focus on the real even part, only. The reason is that this part is sufficient to recover the usual Dirac equation (see Section 4). The only reference we found dealing with other contributions is \cite{Sobczyk} but in a different context than the one involved here (the accent was there put on idempotent matrices).\\
If we follow the realization (7), these four contributions respectively read
\setcounter{equation}{14}
\begin{equation}
\left(
\begin{array}{cc}
N & 0 \\
0 & \widehat{N}
\end{array}
\right), \quad \left(
\begin{array}{cc}
0 & \widehat{N} \\
N & 0
\end{array}
\right), \quad \left(
\begin{array}{cc}
N & 0 \\
0 & -\widehat{N}
\end{array}
\right)
\label{eq: 15}, \quad \left(
\begin{array}{cc}
0 & -\widehat{N} \\
N & 0
\end{array}
\right).
\end{equation}
Here the matrix $N$ is given by
\begin{equation}
N=(a^0+i b^0) I+(a^j+i b^j) \sigma_j,
\label{eq: 16}
\end{equation}
while $\widehat{N}$ refers, in agreement with the usual notation,  to 
\begin{equation}
\widehat{N}=(a^0-i b^0) I-(a^j-i b^j) \sigma_j.
\label{eq: 17}
\end{equation}
This involutive automorphism corresponds to the fact that each of the 3D basis vector has been reversed.\\
The incoming calculations will be really simplified by remembering that
\begin{equation}
N=\left(
\begin{array}{cc}
a & b \\
c & d
\end{array}
\right) \quad \rightarrow \quad \widehat{N}=\left(
\begin{array}{cc}
d^* & -c^* \\
-b^* & a^*
\end{array}
\right), 
\label{eq: 18}
\end{equation}
where * is the usual complex conjugate and $a,b,c,d$ four complex numbers.\\
Let us also mention that the operators in Eqs.~(14b)-(14d) are not chosen at random. In fact, as is well known, $\gamma_0$ is the parity operator $P$ for the Dirac equation while $-i \; I_4$ actually is the chirality operator $\gamma_5$. However we prefer to see this operator as the product of $P$ and the last one $-i \; \gamma_1 \gamma_2 \gamma_3$   which has been identified \cite{DPD2018} as the unitary time-reversal operator $T$ for the Dirac equation.\\
Consequently, the four contributions read
\begin{equation}
E, \quad O=P  E, \quad IE= PT  E, \quad IO=T  E.
\label{eq: 19}
\end{equation}
This highly suggests to associate them to the four connected components of the Lorentz group. That's precisely the aim of the next Section.
\section{The relativistic invariance}
Following the standard STA approach, we define the position vector by
\begin{equation}
X=x^{\mu} \; \gamma_{\mu}=\left(
\begin{array}{cc}
0 & x \\
\widehat{x} & 0
\end{array}
\right), 
\label{eq: 20}
\end{equation}
where $x=x^0 I+x^j \sigma_j, \widehat{x}=x^0 I-x^j \sigma_j$. \\
Let us now consider the transformations that this position vector can undergo. The general relation is
\begin{equation}
X'= M X \widetilde{M}.
\label{eq: 21}
\end{equation}
The notation $\widetilde{M}$ refers to a particular conjugation (called reversal) which consists in totally swapping the order of the vectors ($\gamma_j \gamma_k \; \rightarrow \; \gamma_k \gamma_j,\gamma_j \gamma_k \gamma_l \; \rightarrow \; \gamma_l \gamma_k \gamma_j$ and so on). It has been chosen so because, in general, $M \widetilde{M} ={\rm diag}(\alpha I,\alpha^*  I)$, a property rather similar to the unitarity of transformations. Moreover, when matrices of dimension 2 are concerned, this conjugation is identical to the hermitic conjugation
\begin{equation}
N=\left(
\begin{array}{cc}
a & b \\
c & d
\end{array}
\right) \quad \rightarrow \quad \widetilde{N}=N^{\dagger}=\left(
\begin{array}{cc}
a^* & c^* \\
b^* & d^*
\end{array}
\right). 
\label{eq: 22}
\end{equation}
We will also need a third conjugation combining the first two ones 
\begin{equation}
N=\left(
\begin{array}{cc}
a & b \\
c & d
\end{array}
\right) \quad \rightarrow \quad \bar{N} \equiv \widetilde{(\widehat{N})}=\left(
\begin{array}{cc}
d & -b \\
-c & a
\end{array}
\right). 
\label{eq: 23}
\end{equation}
The transformations can then be of four types as suggested by Eq.~(19)
$$
X'=M_E X \widetilde{M_E}=\left(
\begin{array}{cc}
N & 0 \\
0 & \widehat{N}
\end{array}
\right) \left(
\begin{array}{cc}
0 & x \\
\widehat{x} & 0
\end{array}
\right) \left(
\begin{array}{cc}
\bar{N} & 0 \\
0 & \widetilde{N}
\end{array}
\right)  \rightarrow  x'=N x \widetilde{N},
\eqno{(24a)}
$$
$$
X'=M_O X \widetilde{M_O}=\left(
\begin{array}{cc}
0 & \widehat{N} \\
N & 0
\end{array}
\right) \left(
\begin{array}{cc}
0 & x \\
\widehat{x} & 0
\end{array}
\right) \left(
\begin{array}{cc}
0 & \bar{N} \\
\widetilde{N} & 0
\end{array}
\right)  \rightarrow  x'=\widehat{N} \widehat{x} \bar{N},
\eqno{(24b)}
$$
$$
X'=M_{IE} X \widetilde{M_{IE}}=\left(
\begin{array}{cc}
N & 0 \\
0 & -\widehat{N}
\end{array}
\right) \left(
\begin{array}{cc}
0 & x \\
\widehat{x} & 0
\end{array}
\right) \left(
\begin{array}{cc}
\bar{N} & 0 \\
0 & -\widetilde{N}
\end{array}
\right)  \rightarrow  x'=-N x \widetilde{N},
\eqno{(24c)}
$$
$$
X'=M_{IO} X \widetilde{M_{IO}}=\left(
\begin{array}{cc}
0 & -\widehat{N} \\
N & 0
\end{array}
\right) \left(
\begin{array}{cc}
0 & x \\
\widehat{x} & 0
\end{array}
\right) \left(
\begin{array}{cc}
0 & \bar{N} \\
-\widetilde{N} & 0
\end{array}
\right)  \rightarrow  x'=-\widehat{N} \widehat{x} \bar{N}.
\eqno{(24d)}
$$
\setcounter{equation}{24}
Each of these transformations (24) can be rewritten in a more usual way as
\begin{equation}
x'=R \; x.
\label{eq: 25}
\end{equation}
Besides, in all cases, the three rotations are given by
\begin{equation}
N_{\rm{rotations}}= e^{\frac{i}{2} \theta \; \sigma_j},
\label{eq: 26}
\end{equation}
(we write the same parameter $\theta$ for the simplicity of the notation but it is evidently subtended that each transformation has its own parameter) while the boosts are obtained from
\begin{equation}
N_{\rm{boosts}}= e^{\frac{1}{2} \theta \; \sigma_j}.
\label{eq: 27}
\end{equation}
If we allow the determinant of $N$ to differ from 1 (and it is without any consequence on $R$ whose determinant still is 1), we also have to add a supplementary transformation characterized by
\begin{equation}
N_{\rm{supplementary}}= e^{\frac{i}{2} \theta \; I}.
\label{eq: 28}
\end{equation}
In the case of the transformation (24a), it is the well-known homomorphism between the special linear group ${\rm SL}(2,C)$ (related to the matrix $N$ when we restrict ourselves to the cases where det $N=1$ -a restriction that we will give up in the following-) and the restricted Lorentz group ${\rm SO}^+ (1,3)$ (associated with $R$). It has to be mentioned that the transformations recovered here -namely three rotations and three boosts- are proper and orthochronous (component ${\rm L}_+^{\uparrow}$ of the Lorentz group). In other words, they preserve the orientation of space (a fact that can be seen from det $R=1$) and the direction of time (the element at the intersection of the first row and the first column of $R$ is positive). \\
The explicit form of the transformation matrices coming from the geometric algebra point of view is given by
$$
M_{\rm{rotations}}=e^{-\frac{1}{2} \; \theta \; \gamma_j \gamma_k}, \quad M_{\rm{boosts}}=e^{-\frac{1}{2} \; \theta \; \gamma_0 \gamma_k}, \quad M_{\rm{supplementary}} = e^{\frac{1}{2} \; \theta \; I_4}.
\eqno{(29a)}
$$
At the algebra level, we thus have to consider
$$
J_{jk} \equiv -\frac{1}{2} \; \gamma_j \gamma_k, \quad J_{0j} \equiv -\frac{1}{2} \; \gamma_0 \gamma_j, \quad J \equiv \frac{1}{2} \; I_4.
\eqno{(29b)}
$$
As well known, the six first operators generate the Lorentz algebra ${\rm so}(1,3)$ 
\setcounter{equation}{29}
\begin{equation}
[J_{\alpha \beta}, J_{\lambda \rho}]=g_{\beta \lambda}\; J_{\rho \alpha}+g_{\alpha \rho} \; J_{\lambda \beta}+g_{\beta \rho} \; J_{\alpha \lambda}+g_{\alpha \lambda} \; J_{\beta \rho}, \quad g={\rm diag}(1,-1,-1,-1),
\label{eq: 30}
\end{equation}
while $J$ (still with a vanishing trace) commutes with each of these operators.\\
To find back the other three connected components, the three other cases (24b)-(24d) are necessary. \\
The proper retrochronous transformations (i.e. ${\rm L}_+^{\downarrow}$ for which the time is reversed, a fact that is marked through a change of sign in $R_{00}$) are obtained through Eq.~(24c). Indeed the result (24c) just means that $R$ has been replaced by $-R$. At the level of the rotations, the interpretation is that not only the time but also one of the spatial coordinates has been reversed. The rotations themselves, in each of the planes subtended by two spatial coordinates, are no longer rotations of angle $\theta$ in the clockwise direction but rotation of angle $\pi - \theta$ in the trigonometric direction. At the boosts, all the coordinates have been reversed but the velocity is unchanged. This total inversion will be also found at the additional symmetry: $x'=-x$.\\
The explicit form of the corresponding transformation matrices are obtained from Eq.~(29a) on which $PT=-i \; I_4$ is applied. Obviously, they cannot be written as exponentials anymore. However the relations (30) are still valid if we consider the seven operators (29b) on which $PT$ is applied on the left.\\
The improper orthochronous component (i.e. ${\rm L}_-^{\uparrow}$ for which the orientation of space has been changed or, in other words, the determinant of $R$ is now equal to $-1$) is obtained from the odd transformation (24b). In fact, with respect to the original matrix $R$, assigned to the proper orthochronous transformations, only the first row is unchanged. The other rows have changed sign. In what concerns the rotations, the time is unchanged while a spatial coordinate is reversed. Once again, in each of the plane determined by the two other spatial coordinates, there is no longer a rotation of $\theta$ in the clockwise direction but a rotation of $\pi - \theta$ in the trigonometric sense. For the boosts, we notice that the time remains unchanged while the spatial coordinates have been inversed and so is the velocity.\\
The explicit form of the corresponding transformation matrices is obtained from Eqs.~(29) on which $P=\gamma_0$ is applied. \\
Let us also mention that the supplementary transformation is nothing but this $P$ operation. The corresponding matrix $R$ gets the well-known form ${\rm diag}(1,-1,-1,-1)$.\\
The improper retrochronous transformations (${\rm L}_-^{\downarrow}$) correspond to Eq.~(24d). The situation is the reverse of the previous one: only the first row of the matrix $R$ has changed sign. So the rotations are the ones of the proper orthochronous transformations and only time has been reversed. The boosts are, as in the previous context, characterized by an inversion of the velocity due, in this case, to the reversal of time (the spatial coordinates remain unchanged with respect to the proper orthochronous boosts). \\
As expected, the explicit form of the corresponding transformations is obtained from Eqs.~(29) on which $T=-i \;\gamma_1 \gamma_2 \gamma_3$ is applied. It is thus not surprising that the supplementary transformation reduces to this unitary time-reversal transformation and the corresponding $R$ is ${\rm diag}(-1,1,1,1)$.\\
Let us end this Section by mentioning that ${\rm gl}(2,C)$ is usually recognized as the Lie algebra associated with the even part of the STA. It is due to the presence of the matrix $N$ on which no constraint is required. In particular, the determinant ($=ad-bc$) can take any complex value, say $r e^{i\; \theta}$. We have focused here on the case where $r=1$.  Nevertheless, there remains an additional parameter (with respect to the Lorentz group) in the presence of the angle, which is associated with this operator $J$.\\
If we consider the odd part in addition (as it is the case when we deal with the improper transformations), we speak about a Lie superalgebra, that is precisely ${\rm gl}(2|2)$. This superalgebra is characterized by the relations \cite{Xiang2003}
\begin{equation}
[E_{pq}, E_{rs}]=\delta_{qr} \; E_{ps}- (-1)^{((p)+(q))((r)+(s))} \delta_{ps} E_{rq},
\label{eq: 31}
\end{equation}
with the following rules : if $p+q$ and $r+s$ are odd numbers, the Lie bracket is an anticommutator while it reduces to a commutator in all other cases.\\
To be convinced of this link, we realize the Cartan subalgebra generators with
$$
H_1 =E_{11} +E_{33} =\frac{1}{2}I-\frac{i}{2} I_4, \; H_2 =E_{22} +E_{33} =\frac{1}{2}I-\frac{i}{2} \gamma_1 \gamma_2,
\eqno{(32a)}
$$
$$
H_3 =E_{22} +E_{44} =\frac{1}{2}I+\frac{i}{2} I_4,
\eqno{(32b)}
$$
$$
H_4=E_{11}+E_{22}-E_{33}-E_{44}+\alpha \; (E_{11}+E_{22}+E_{33}+E_{44})=-\gamma_0 \gamma_3 + \alpha \; I.
\eqno{(32c)}
$$
We complete the identification by realizing the simple raising operators (two even operators and one odd operator) as
$$
E_{13}=-\frac{i}{4} \; \gamma_2 \gamma_3+\frac{1}{4} \; \gamma_0 \gamma_1+\frac{1}{4} \; \gamma_3 \gamma_1+\frac{i}{4} \; \gamma_0 \gamma_2,
\eqno{(33a)}
$$
$$
E_{24}=-\frac{i}{4} \; \gamma_2 \gamma_3-\frac{1}{4} \; \gamma_0 \gamma_1-\frac{1}{4} \; \gamma_3 \gamma_1+\frac{i}{4} \; \gamma_0 \gamma_2,
\eqno{(33b)}
$$
$$
E_{32}=-\frac{1}{4} \; \gamma_0 +\frac{1}{4} \;  \gamma_3+\frac{i}{4} \; \gamma_0 \gamma_1 \gamma_2-\frac{i}{4} \; \gamma_1 \gamma_2 \gamma_3,
\eqno{(33c)}
$$
together with their lowering counterparts
$$
E_{31}=E_{13}^{\dagger}, \quad E_{42}=E_{24}^{\dagger}, \quad E_{23}=E_{32}^{\dagger}.
\eqno{(33d)}
$$
\setcounter{equation}{33}
Finally, we add the following odd generators
\begin{equation}
E_{12}=E_{13} \; E_{32}, \quad E_{34}=E_{32} \; E_{24}, \quad E_{14}=E_{12} \; E_{24},
\label{eq: 34}
\end{equation}
where the product involved here is the geometric one.\\
Let us now turn our attention to the Dirac equation.
\section{The Dirac equation with positive and negative masses}
A remark to begin with: when Dirac established his famous equation, in the free case, he only imposed conditions on the squares of the matrices (see Eq.~(6) of \cite{Dirac1928}) leaving the field open to two possibilities of signs before the mass term. On page 615 of his paper, he specifies that he marks one possible choice to satisfy these conditions.\\
So, at the start, the possibility of having negative masses was already present.\\
Now, let us consider the Dirac equation when the electron is in interaction with an electromagnetic field
\begin{equation}
\biggl(\gamma_{\mu}(\partial_{\mu}+i\; e \; A_{\mu})+i \; m \biggr)\; |\phi\rangle_+ = 0.
\label{eq: 35}
\end{equation}
With the realization (7) and $|\phi\rangle_+  \equiv \left(
\begin{array}{c}
\xi_+ \\
\eta_+
\end{array}
\right)  $, this equation can be written as
$$
\left\{
\begin{array}{ll}
(\Delta+i \; e \; A) \; \eta_++i \; m \; \xi_+ =0\\
(\widehat{\Delta}+i \; e \; \widehat{A}) \; \xi_++i \; m \; \eta_+ =0
\end{array}
\right.
\eqno{(36)}
$$
where the usual gradient $\Delta$ issued from STA has been used as well as its conjugate
\setcounter{equation}{36}
$\widehat{\Delta}$
\begin{equation}
\Delta \equiv \partial_0-\sigma^j \; \partial_j, \quad \widehat{\Delta} \equiv \partial_0+\sigma^j \; \partial_j.
\label{eq: 37}
\end{equation}
To be coherent, an analogue of the electromagnetic field has also been defined
\begin{equation}
A \equiv A_0-\sigma^j \; A_j, \quad \widehat{A} \equiv A_0+\sigma^j \; A_j.
\label{eq: 38}
\end{equation}
By taking the complex conjugate of the system (36) and then multiplying it, on the right, by $-i \sigma_2$, we obtain
$$
\left\{
\begin{array}{ll}
(\widehat{\Delta}-i \; e \; \widehat{A}) \; (-i\sigma_2 \eta^*_+)-i \; m \; (-i\sigma_2 \xi^*_+) =0 \nonumber\\
(\Delta-i \; e \; A) \; (-i\sigma_2 \xi^*_+)-i \; m \; (-i\sigma_2 \eta^*_+)=0
\end{array}
\right.
\eqno{(39)}
$$
\setcounter{equation}{39}
Here we have made use of
\begin{equation}
-i \sigma_2 \; \widehat{\Delta}^*=\Delta \; (-i \sigma_2), \quad-i \sigma_2 \; \widehat{A}^*=A \; (-i \sigma_2). 
\label{eq: 40}
\end{equation}
We are now ready to highlight four types of matter by looking closely at the systems (36) and (39).
Indeed, the system (36) shows us that if the electron of positive matter
$$
| \phi\rangle_{e,+}=\left(
\begin{array}{c}
\xi_+ \\
\eta_+
\end{array}
\right) \equiv \left(
\begin{array}{c}
\phi_1 \\
\phi_2 \\
\phi_3 \\
\phi_4
\end{array}
\right),
\eqno{(41a)}
$$
is the usual solution, there is an additional solution corresponding to a change of sign in the last term and thus to an electron of negative mass. It is given by
$$
| \phi\rangle_{e,-}=\left(
\begin{array}{c}
\xi_+ \\
-\eta_+
\end{array}
\right) \equiv \left(
\begin{array}{c}
\phi_1 \\
\phi_2 \\
-\phi_3 \\
-\phi_4
\end{array}
\right).
\eqno{(41b)}
$$
In what concerns the system (39), it tells us that there is a solution associated with a positron of positive mass
$$
| \phi\rangle_{p,+}=\left(
\begin{array}{c}
-i \sigma_2 \eta^*_+ \\
i \sigma_2 \;\xi^*_+
\end{array}
\right) \equiv \left(
\begin{array}{c}
-\phi_4^* \\
\phi_3^* \\
\phi_2^* \\
-\phi_1^*
\end{array}
\right),
\eqno{(41c)}
$$
as well as a solution corresponding to a positron of negative mass
$$
| \phi\rangle_{p,-}=\left(
\begin{array}{c}
-i \sigma_2 \eta^*_+ \\
-i \sigma_2 \;\xi^*_+
\end{array}
\right) \equiv \left(
\begin{array}{c}
-\phi_4^* \\
\phi_3^* \\
-\phi_2^* \\
\phi_1^*
\end{array}
\right).
\eqno{(41d)}
$$
We notice easily that
\setcounter{equation}{41}
\begin{equation}
| \phi\rangle_{p,+}=i \; \gamma^2 \;| \phi\rangle^*_{e,+}, \quad | \phi\rangle_{p,-}=-i \; \gamma^2 \;| \phi\rangle^*_{e,-}. 
\label{eq: 42}
\end{equation}
The first of these two relations is what is expected from the well-known charge conjugation operator. A change of sign with no consequence (a ket being defined up to a phase factor) is introduced within the negative masses counterpart.\\
We also see that
\begin{equation}
| \phi\rangle_{e,-}=-i\; I_4 \;| \phi\rangle_{e,+}, \quad | \phi\rangle_{p,-}=-i\; I_4 \; | \phi\rangle_{e,-}. 
\label{eq: 43}
\end{equation}
Now, the matrix $-i\; I_4$ , belonging to the imaginary extension of the STA, is nothing else than the chirality operator $\gamma_5$. 
\section{The Dirac-Hestenes equation with positive and negative masses}
We owe it to Hestenes to have proposed the counterpart of the Dirac equation in the context of geometric algebra. This (real) equation is
\begin{equation}
\gamma_{\mu} \; \biggl(\partial_{\mu} \phi \; \gamma_2 \gamma_1-e \; A_{\mu}\; \phi \biggr)=m \; \phi \; \gamma_0.
\label{eq: 44}
\end{equation}
It can also be proposed on the form
$$
\left\{
\begin{array}{ll}
-i \Delta \; \widehat{\Phi} \; \sigma_3+e\; A \; \widehat{\Phi}+m\; \Phi=0 \\
-i \widehat{\Delta} \; \Phi \; \sigma_3+e\; \widehat{A} \; \Phi+m\; \widehat{\Phi}=0 
\end{array}
\right.
\eqno{(45)}
$$
\setcounter{equation}{45}
with $\phi=\left( \begin{array}{cc}
\Phi & 0 \\
0 & \widehat{\Phi} 
\end{array}
\right)$.\\
It is then usually said that the whole Dirac formalism is contained in the first  2 by 2 equation of Eq.~(45) as the second one is simply its conjugate in the sense $\sigma_j \rightarrow -\sigma_j$.\\
Hestenes' demonstration of his equation is as follows: he proposed to consider a bispineur $|u \rangle$ such that
\begin{equation}
\langle u | u \rangle =1, \quad \gamma_0 | u \rangle = | u \rangle, \quad \gamma_2 \gamma_1 | u \rangle = i | u \rangle.
\label{eq: 46}
\end{equation}
It is then sufficient to put
\begin{equation}
| \phi \rangle_+ =\phi \; |u \rangle,
\label{eq: 47}
\end{equation}
to recover Eq.~(35). \\
If we are interested in the precise form of the ket $|u\rangle $ satisfying Eq.~(46) with the realization (7), we obtain
\begin{equation}
| u \rangle =\frac{1}{\sqrt{2}} \; \left( \begin{array}{c}
1 \\
0 \\
1 \\
0  
\end{array}
\right),
\label{eq: 48}
\end{equation}
up to a constant $C$ such that $|C|=1$.\\
Eq.~(45) and Eqs.~(47)-(48) lead to
\begin{equation}
\phi \equiv \phi_{+1} = \sqrt{2} \; \left( \begin{array}{cccc}
\phi_1 & -\phi^*_4 & 0 & 0\\
\phi_2 & \phi^*_3 & 0 & 0 \\
0 & 0 & \phi_3 & -\phi^*_2 \\
0 & 0 & \phi_4 & \phi^*_1
\end{array}
\right)= \left( \begin{array}{cc}
\Phi & 0 \\
0 & \widehat{\Phi} 
\end{array}
\right).
\label{eq: 49}
\end{equation}
At this stage, we have to make a few remarks.\\
First, the STA approach allows us to see unusual facts. By this we mean that the first block on the diagonal contains all the information of the traditional theory. But what do we see there? The simultaneous consideration of the electron of positive mass and of what seems to be, at first sight, the positron of positive mass (cf. the first two components of Eq.~(41a) and Eq.~(41c)). But if we look at it more closely and consider the second block on the diagonal, we see that, if we confirm the presence of the electron with positive mass, the antiparticle is, on the other hand, characterized by a negative mass (cf. Eq.~(41d))!\\
Second, this combination of positive matter-negative antimatter is typical to have considered the real even part of the STA only. Indeed, Eq.~(49) is of type $E$ in Eq.~(15)).\\
So the question arises naturally: is it possible to put in evidence the other solutions Eq.~(41c) (we restrict ourselves to even matrices)?\\
For this purpose, we modify Eq.~(47) and consider, instead
\begin{equation}
| \phi \rangle_- = -i\; I_4 \; | u \rangle.
\label{eq: 50}
\end{equation}
The new matrix $-i\; I_4 \; \phi$ is then of $IE$-type and writes explicitly
\begin{equation}
-i \;I_4 \;\phi \equiv \phi_{-1} = \sqrt{2} \; \left( \begin{array}{cccc}
\phi_1 & -\phi^*_4 & 0 & 0\\
\phi_2 & \phi^*_3 & 0 & 0 \\
0 & 0 & -\phi_3 & \phi^*_2 \\
0 & 0 & -\phi_4 & -\phi^*_1
\end{array}
\right)= \left( \begin{array}{cc}
\Phi & 0 \\
0 & -\widehat{\Phi} 
\end{array}
\right).
\label{eq: 51}
\end{equation}
So we now have the combination of negative matter-positive antimatter and the Dirac-Hestenes equation is slightly modified to write
\begin{equation}
\gamma^{\mu} \; \biggl( \partial_{\mu} \left(-i \; I_4 \;\phi \right) \; \gamma_2 \gamma_1-e \; A_{\mu}\; \left(-i \;I_4 \;\phi \right)\biggr)=-m \; \left(-i\; I_4 \;\phi \right) \; \gamma_0.
\label{eq: 52}
\end{equation}
Compared to the original one (44), we notice a change of sign in what concerns the mass term. The four types of matter (antimatter) can thus be embedded in a system made up of  Eqs.~(44) and (52).\\
Let us now have a look on the transformations of the wave functions.
\section{Relativistic invariance of the Dirac equation with positive and negative masses}
We concentrate here on the free case since it is implied that the transformations of the derivatives and components of the electromagnetic field are identical in form.\\
The question is the following: how can we characterize the transformation $|\phi'\rangle_{\pm}$   on $|\phi\rangle_{\pm}$   such that
$$
(\gamma^{\mu} \; \partial_{\mu} \pm i \; m) \; |\phi\rangle_{\pm}=0,
\eqno{(53a)}
$$
transforms as
$$
(\gamma^{\mu} \; \partial'_{\mu} \pm i \; m') \; |\phi'\rangle_{\pm}=0,
\eqno{(53b)}
$$
if $\partial'_{\mu}$ is specified according to one of the transformations (24)? \\
This question has already been answered for the transformation (24a) in \cite{Daviau2013}.\\
We resume the results by presenting them from a point of view that can be applied to other transformations (24).\\
First, we notice that the transformation (24a), $x'=N \; x \widetilde{N}$, implies
\setcounter{equation}{53}
\begin{equation}
\Delta = \bar{N} \; \Delta' \; \hat{N},
\label{eq: 54}
\end{equation}
where $\Delta$ has been defined in Eq.~(37).\\
Second, we introduce the transformation (54) in Eq.~(53a) and ask the result to be equivalent to
\begin{equation}
M_E^{-1} \; (\gamma^{\mu} \; \partial'_{\mu} \pm i \; m') \; M_E \; | \phi >_{\pm}.
\label{eq: 55}
\end{equation}
If so, by applying $M_E$  on the left of Eq.~(55), this implies, at least for the six Lorentz transformations, that
\begin{equation}
m'=m, \; | \phi' \rangle_{\pm}=M_E \; | \phi \rangle_{\pm}.
\label{eq: 56}
\end{equation}
The key is to ensure that
\begin{equation}
\gamma_{\mu} \; M_E \; \partial'_{\mu}=M_E \; \gamma_{\mu} \; \partial_{\mu},
\label{eq: 57}
\end{equation}
with $\partial'_{\mu}$ being given by Eq.~(54). As easily checked, this is satisfied with $M_E$ given by Eq.~(29a).\\
As an example, let us consider the rotation 
$$
\left\{
\begin{array}{ll}
x'^0 = x^0  \\
x'^1 = x^1\\
x'^2=\cos{\theta} \; x^2+\sin{\theta} \; x^3 \\
x'^3=-\sin{\theta} \; x^2+\cos{\theta} \; x^3
\end{array}
\right.
\eqno{(58)}
$$\setcounter{equation}{58}
The constraint (57) leads then to
$$
\left\{
\begin{array}{ll}
\gamma^0 \; M_E = M_E \; \gamma^0, \; \gamma^1 \; M_E = M_E \; \gamma^1  \\
\gamma^2 \; M_E = M_E \; (\gamma^2 \cos{\theta}+\gamma^3 \sin{\theta}), \; \gamma^3 \; M_E = M_E \; (-\gamma^2 \sin{\theta}+\gamma^3 \cos{\theta}).
\end{array}
\right.
\eqno{(59)}
$$
\setcounter{equation}{59}
These relations are indeed satisfied with
\begin{equation}
M_E = \cos{\frac{\theta}{2}} \; I - \sin{\frac{\theta}{2}} \; \gamma^2 \gamma^3.
\label{eq: 60}
\end{equation}
The same results hold true for each of the six Lorentz transformations. \\
The case of the supplementary operator $M_E=e^{\frac{1}{2} \; \theta \;  I_4 }$ is a bit trickier. Indeed, in this case, as $x'=x,\Delta'=\Delta$, we have two possibilities: either $m'=m$ and $|\phi'\rangle_{\pm}=|\phi\rangle_{\pm}$ or $m'\neq m$ and $|\phi'\rangle_{\pm}=M_E \; |\phi\rangle_{\pm}$.  The first possibility is the only one allowed by the traditional way of seeing while the second possibility is typical of the geometric algebra point of view which is richer compared to the usual one because the subtended group ${\rm GL}(2,C)$  has more parameters than the Lorentz one.\\
It seems to us much more logical to favor this second possibility for reasons of consistency as, for the Lorentz transformations, we precisely have $|\phi'\rangle_{\pm}=M_E \; |\phi\rangle_{\pm}$ each time.   We therefore opt for, in all cases 
\begin{equation}
|\phi'\rangle_{\pm}=M_E \; |\phi\rangle_{\pm}.
\label{eq: 61}
\end{equation}
We also notice that, for $M_E=e^{\frac{1}{2} \theta \; I_4 }$ , we have
\begin{equation}
\gamma^{\mu} \; M_E=M_E^{-1}\gamma^{\mu},
\label{eq: 62}
\end{equation}
and finally obtain
\begin{equation}
(\gamma^{\mu} \; \partial'_{\mu} \pm i m') \; |\phi' \rangle_{\pm}=\pm i \;(-M_E^{-1} \; m+m' \; M_E) \; |\phi \rangle_{\pm}. 
\label{eq: 63}
\end{equation}
So, the invariance leads to
\begin{equation}
m' = m \; \left( \begin{array}{cc}
e^{-i \; \theta} &  0\\
0 & e^{i \; \theta}
\end{array}
\right).
\label{eq: 64}
\end{equation}
What this tells us is that the mass itself undergoes a transformation, which consists of a rotation in the complex plane. For $\theta=\pi$, these rotations lead to an inversion of the sign of the mass. \\
As noticed in \cite{Daviau2013}, the relations (62) lead to different transformation laws for each spinor of $|\phi\rangle_{\pm} $\begin{equation}
| \xi' \rangle_{\pm}=N \; | \xi \rangle_{\pm} \; ; \; | \eta' \rangle_{\pm}=\widehat{N} \; | \eta \rangle_{\pm}.
\label{eq: 65}
\end{equation}  
In what concerns the version coming from the geometric algebra, it is easy to convince ourselves that
\begin{equation}
\phi'_{\epsilon}=M_E \; \phi_{\epsilon} \; ; \; \epsilon=\pm 1.
\label{eq: 66}
\end{equation}
Let us now turn, in a similar way, to the other cases (24c), (24b) and (24d).
First, we consider the proper retrochronous transformations. At the start, we observe a simple change of sign through Eq.~(24c) and
\begin{equation}
\Delta = -\bar{N} \; \Delta' \; \hat{N}.
\label{eq: 67}
\end{equation}
This leads to anticommutation relations where there were commutation relations and vice versa. In particular, Eq.~(59) now reads 
$$
\left\{
\begin{array}{ll}
\gamma^0 \; M_{IE} =- M_{IE} \; \gamma^0, \; \gamma^1 \; M_{IE} = -M_{IE} \; \gamma^1  \\
\gamma^2 \; M_{IE} =- M_{IE} \; (\gamma^2 \cos{\theta}+\gamma^3 \sin{\theta}), \; \gamma^3 \; M_{IE} = M_{IE} \; (\gamma^2 \sin{\theta}-\gamma^3 \cos{\theta})
\end{array}
\right.
\eqno{(68)}
$$
\setcounter{equation}{68}
and is satisfied with
\begin{equation}
M_{IE}=-i\; \sin{\frac{\theta}{2} \; \gamma_0 \gamma_1}-i\; \cos{\frac{\theta}{2}\; \gamma_0 \gamma_1 \gamma_2 \gamma_3}=-i \; I_4 \; M_E,
\label{eq: 69}
\end{equation}
as expected.\\
This change of sign is also recovered for the supplementary symmetry in the sense that we now have
\begin{equation}
\gamma^{\mu} \; M_{IE}=-M^{-1}_{IE} \; \gamma_{\mu}.
\label{eq: 70}
\end{equation}
Despite this change of sign, the relation (64) still holds true.\\
In what concerns the results (65) and (66), they become respectively
\begin{equation}
| \xi' \rangle_{\pm}=N \; | \xi \rangle_{\pm} \; ; \; | \eta' \rangle_{\pm}=-\hat{N} \; | \eta \rangle_{\pm},
\label{eq: 71}
\end{equation}  
and
\begin{equation}
\phi'_{\epsilon}=i \; M_{IE} \; \phi_{\epsilon} \; \gamma_1 \gamma_2.
\label{eq: 72}
\end{equation}
We now come to the improper orthochronous transformations characterized by Eq.~(24b). The transformation law on the derivatives is given by
\begin{equation}
\Delta = \widetilde{N} \; \hat{\Delta'} \; N.
\label{eq: 73}
\end{equation}
Each constraint on $M_O$ is affected by sign changes. For instance, Eq.~(59) now reads
$$
\left\{
\begin{array}{ll}
\gamma^0 \; M_{O} = M_{O} \; \gamma^0, \; \gamma^1 \; M_{O} = -M_{O} \; \gamma^1  \\
\gamma^2 \; M_{O} =- M_{O} \; (\gamma^2 \cos{\theta}+\gamma^3 \sin{\theta}), \; \gamma^3 \; M_{O} = M_{O} \; (\gamma^2 \sin{\theta}-\gamma^3 \cos{\theta})
\end{array}
\right.
\eqno{(74)}
$$
\setcounter{equation}{74}
and is satisfied with
\begin{equation}
M_O=\cos{\frac{\theta}{2}} \; \gamma_0-\sin{\frac{\theta}{2}} \; \gamma_0 \gamma_2 \gamma_3= \gamma_0 \; M_E.
\label{eq: 75}
\end{equation}
In what concerns the supplementary symmetry, we have to distinguish the time contribution from the spatial ones 
$$
\gamma^0 \; M_O=\gamma^0  \left(\cos{\frac{\theta}{2}} \; \gamma_0+\sin{\frac{\theta}{2}} \; \gamma_1 \gamma_2 \gamma_3 \right)=\left(\cos{\frac{\theta}{2}} \; \gamma_0-\sin{\frac{\theta}{2}} \; \gamma_1 \gamma_2 \gamma_3 \right)  \gamma^0,
\eqno{(76a)}
$$
$$
\gamma^j \; M_O=\gamma^j  \left(\cos{\frac{\theta}{2}} \; \gamma_0+\sin{\frac{\theta}{2}} \; \gamma_1 \gamma_2 \gamma_3 \right)=-\left(\cos{\frac{\theta}{2}} \; \gamma_0-\sin{\frac{\theta}{2}} \; \gamma_1 \gamma_2 \gamma_3 \right)  \gamma^j.
\eqno{(76b)}
$$
\setcounter{equation}{76}
in order to obtain the transformation law (64) on mass.\\
The results (65)-(66) are now
\begin{equation}
| \xi' \rangle_{\pm}=\hat{N} \; | \eta \rangle_{\pm} \; ; \; | \eta' \rangle_{\pm}=N \; | \xi \rangle_{\pm}.
\label{eq: 77}
\end{equation}  
We conclude this case by mentioning that
\begin{equation}
\phi'_{\epsilon}
= \epsilon \; M_O \; \phi_{\epsilon} \; \gamma_0.
\label{eq: 78}
\end{equation}
Let us now turn to the ultimate case, the one of improper retrochronous transformations (24d). It is extremely close to the previous case. Eqs.~(73) and (74) just get a minus sign in front of the second member of each relation. Consequently, we obtain
\begin{equation}
M_{IO}=-i \; \cos{\frac{\theta}{2}} \; \gamma_1 \gamma_2 \gamma_3-i \; \sin{\frac{\theta}{2}} \; \gamma_1 = -i \; \gamma_1 \gamma_2 \gamma_3 \; M_E.
\label{eq: 79}
\end{equation}
Relations (76) become
$$
\gamma^0 \; M_{IO}=\gamma^0  \left(i\; \cos{\frac{\theta}{2}} \; \gamma_1 \gamma_2 \gamma_3-i \; \sin{\frac{\theta}{2}} \; \gamma_0 \right)=\left(-i \; \cos{\frac{\theta}{2}} \; \gamma_1 \gamma_2 \gamma_3-i\; \sin{\frac{\theta}{2}} \; \gamma_0 \right)  \gamma^0,
\eqno{(80a)}
$$
$$
\gamma^j \; M_{IO}=\gamma^j  \left(i\; \cos{\frac{\theta}{2}} \; \gamma_1 \gamma_2 \gamma_3-i \; \sin{\frac{\theta}{2}} \; \gamma_0 \right)=\left(i \; \cos{\frac{\theta}{2}} \; \gamma_1 \gamma_2 \gamma_3+i\; \sin{\frac{\theta}{2}} \; \gamma_0 \right)  \gamma^j,
\eqno{(80b)}
$$
leading once again to Eq.~(64).\\
The information is complete if we add
\setcounter{equation}{80}
\begin{equation}
| \xi' \rangle_{\pm}=\hat{N} \; | \eta \rangle_{\pm} \; ; \; | \eta' \rangle_{\pm}=-N \; | \xi \rangle_{\pm},
\label{eq: 81}
\end{equation}  
and
\begin{equation}
\phi'_{\epsilon}
= \epsilon \; M_{IO} \; \phi_{\epsilon} \; \gamma_0 \gamma_1 \gamma_2.
\label{eq: 82}
\end{equation}

\section{Conclusion}
Usually, when one study the spacetime algebra, one focus to its even part, isomorphic to the space algebra. This is because this even part allows to find the whole content of the Dirac equation. The relativistic invariance is then that of the restricted Lorentz group.\\
We wanted to extend these considerations here by proposing three other types of elements of this spacetime algebra, i.e. the odd and the imaginary parts. This allowed us to highlight all connected components of the Lorentz group.\\
We then linked these developments to the Dirac equation, with the two signs of mass and confirmed that the four propositions were associated with four types of matter, of positive or negative mass, of positive or negative charge.\\
Each time, the Lorentz invariance was confirmed but also completed by a new transformation: a rotation in the complex plane concerning the mass and thus allowing the inversion of the sign of this mass.\\
This supplementary transformation is due to the presence of the unit pseudo-scalar of the STA. We are convinced that this pseudo-scalar unit could bring other interesting developments. For example, we know that the mass term of the usual Dirac Lagrangian, $m \bar{\phi} \phi$, is purely scalar. But what happens to this mass term in its geometric algebra version? We just have to take again Eq.~(49) and Eq.~(51) simultaneously on the form
\begin{equation}
\phi_{\epsilon}=\left( \begin{array}{cc}
\Phi & 0 \\
0 & \epsilon \widehat{\Phi} 
\end{array}
\right),
\label{eq: 83}
\end{equation}
as well as
\begin{equation}
\bar{\phi_{\epsilon}}=\left( \begin{array}{cc}
\bar{\Phi} & 0 \\
0 & \epsilon \widetilde{\Phi} 
\end{array}
\right),
\label{eq: 84}
\end{equation}
to see that
\begin{equation}
m \; \bar{\phi_{\epsilon}} \; \phi_{\epsilon}=\left( \begin{array}{cc}
(c1+i c2) I & 0 \\
0 & (c1-i c2) I 
\end{array}
\right), 
\label{eq: 85}
\end{equation}
with
$$
\left\{
\begin{array}{ll}
c1 \equiv \phi_1 \; \phi^*_3+\phi^*_1 \; \phi_3+\phi_2 \; \phi^*_4+\phi_2^* \; \phi_4  \\
c2 \equiv i(-\phi_1 \; \phi^*_3+\phi^*_1 \; \phi_3-\phi_2 \; \phi^*_4+\phi_2^* \; \phi_4)
\end{array}
\right.
\eqno{(86)}
$$
What is remarkable in Eq.~(85) is the combination of both scalar and pseudo-scalar contributions
\setcounter{equation}{86}
\begin{equation}
\label{eq: 87}
m \; \bar{\phi_{\epsilon}} \; \phi_{\epsilon}=c1 + c2 \; I_4,
\end{equation}
a fact that was absent from traditional considerations since
$$m \bar{\phi} \phi=m \; \langle\phi|\gamma_0 |\phi\rangle =m \; c1. \eqno{(88)}$$
We believe that a mass acquisition process "\`a la Higgs" applied to Eq.~(85) could bring interesting developments and this is what we will work on in the future.

\section{Acknowledgments}
Warm thanks are due to the members of the Support au Mod\`ele Cosmologique Janus for their financial support.

\end{document}